\documentclass[prd,preprintnumbers,nofootinbib,superscriptaddress,tightenlines,onecolumn,notitlepage]{revtex4-1}

\usepackage[utf8]{inputenc}
\usepackage{newtxtext}
\usepackage[libertine]{newtxmath}
\usepackage[main=english]{babel}
\usepackage{microtype}

\usepackage{graphicx}
\usepackage[usenames,dvipsnames]{xcolor}
\graphicspath{{figures/},{gallery/}}

\usepackage{hyperref}
\hypersetup{colorlinks=true,citecolor=blue,linkcolor=blue,urlcolor=blue}

\usepackage{diagbox}
\usepackage{booktabs}

\usepackage{amsmath,amsfonts}
\usepackage{mathrsfs}
\usepackage{slashed}
\usepackage{bm}

\usepackage[draft,commentmarkup=uwave]{changes} 
\definechangesauthor[color=orange!76!black]{Adam}
\definechangesauthor[color=blue!50!black]{Ian}
\definechangesauthor[color=red]{Eric}
\definechangesauthor[color=blue]{rev}  

\usepackage{outlines}
\usepackage{enumitem}
\setenumerate[1]{label=\Roman*.}
\setenumerate[2]{label=\Alph*.}
\setenumerate[3]{label=\roman*.}
\setenumerate[4]{label=\alph*.}

\newcommand{\eref}[1]{Eq.~(\ref{e.#1})}

\newcommand{\erefstwo}[2]{Eqs.~(\ref{e.#1})~and~(\ref{e.#2})}
\newcommand{\fref}[1]{Fig.~\ref{f.#1}}

\newcommand{\sref}[1]{Sec.~\ref{s.#1}}

\usepackage[compact]{titlesec}

\newcommand*\dd{\mathop{}\!\mathrm{d}}
\newcommand{\sh}[1]{\slashed{#1}}

\def\lf{\left}
\def\rg{\right}

\newcommand{\ph}[1]{\phantom{#1}}

\begin{document}

\title{Shedding light on shadow generalized parton distributions}

\newcommand*{\ANL}{Physics Division, Argonne National Laboratory, Lemont, Illinois 60439, USA}\affiliation{\ANL}
\newcommand*{\PSU}{Division of Science, Penn State University Berks, Reading, Pennsylvania 19610, USA}\affiliation{\PSU}
\newcommand*{\UW}{Department of Physics, University of Washington, Seattle, Washington 98195, USA}\affiliation{\UW}
\newcommand*{\TU}{Department of Physics, SERC, Temple University, Philadelphia, Pennsylvania 19122, USA}\affiliation{\TU}
\newcommand*{\JLAB}{Jefferson Lab, Newport News, Virginia 23606, USA}\affiliation{\JLAB}
\newcommand*{\ADL}{CSSM and CDMPP, Department of Physics, University of Adelaide, Adelaide 5000, Australia}\affiliation{\ADL}

\author{Eric~Moffat}
\email{emoffat@anl.gov}
\affiliation{\ANL}
\affiliation{\PSU}
\author{Adam Freese}
\affiliation{\UW}
\author{Ian~Clo\"et}
\affiliation{\ANL}
\author{Thomas~Donohoe}
\affiliation{\ANL}
\author{Leonard~Gamberg}
\affiliation{\PSU}
\author{W.~Melnitchouk}
\affiliation{\JLAB}
\affiliation{\ADL}
\author{Andreas~Metz}
\affiliation{\TU}
\author{Alexei~Prokudin}
\affiliation{\PSU}
\affiliation{\JLAB}
\author{Nobuo~Sato}
\affiliation{\JLAB}

\begin{abstract}
The feasibility of extracting generalized parton distributions (GPDs) from deeply-virtual Compton scattering (DVCS) data has recently been questioned because of the existence of an infinite set of so-called ``shadow GPDs'' (SGPDs). These SGPDs depend on the process and manifest as multiple solutions (at a fixed scale $Q^2$) to the inverse problem that needs to be solved to infer GPDs from DVCS data. 
SGPDs therefore pose a significant challenge for extracting GPDs from DVCS data. With this motivation we study the extent to which QCD evolution can provide constraints on SGPDs. This is possible because the known classes of SGPDs begin to contribute to observables after evolution, 
and can then be constrained (at the input scale $Q^2_0$) by data that has a finite $Q^2$ range. The impact that SGPDs could have on determining the total angular momentum, pressure and sheer force distributions, and tomography is also discussed. Our key finding is that scale evolution, coupled with data over a wide range of skewness $\xi$ and $Q^2$, can constrain the class of SGPDs that we studied and potentially make possible the extraction of GPDs from DVCS data over a limited range in the GPD variables. 
\end{abstract}

\maketitle
\section{INTRODUCTION}
The three-dimensional imaging of the quarks and gluons (collectively called partons) confined inside hadrons and nuclei is a key motivation for several particle accelerator facilities around the world, including Jefferson Lab~\cite{Dudek:2012vr,Arrington:2021alx}, COMPASS at CERN~\cite{Sandacz:2018kdx}, J-PARC~\cite{Kumano:2022cje}, and the forthcoming Electron-Ion Collider (EIC)~\cite{Accardi:2012qut,AbdulKhalek:2021gbh}. The spatial imaging of partons is provided through Fourier transforms of generalized parton distributions (GPDs). GPDs are formally defined by bilocal light cone correlators of quark and gluon fields~\cite{Dittes:1988xz, Muller:1994ses, Ji:1996ek, Ji:1996nm, Radyushkin:1996nd, Radyushkin:1996ru} and contain a wealth of information on the momentum and spin distributions of the partons (for some comprehensive reviews see Refs.~\cite{Diehl:2003ny,Belitsky:2005qn,Boffi:2007yc,Kumericki:2016ehc,Mezrag:2022pqk}). For example, the first and second Mellin moments of the leading twist spin-independent GPDs give the electromagnetic and energy-momentum tensor (or gravitational) form factors, respectively, and in the forward limit a subset of the GPDs reduce to the familiar one-dimensional collinear parton distribution functions (PDFs). 

The archetypal process for inferring information on GPDs is deeply-virtual Compton scattering (DVCS)~\cite{Ji:1996nm,Belitsky:2001ns}, which contributes to reactions like $e\,p \to e'\,p'\,\gamma$. The differential cross section for DVCS can be expressed in terms of complex-valued Compton form factors (CFFs) that characterize the target hadron or nucleus. In kinematic domains where quantum chromodynamics (QCD) factorization theorems apply, these CFFs can be expressed as convolutions of complex-valued hard-scattering coefficient functions ($C$) with the real-valued GPDs. Inferring GPDs from DVCS data therefore entails an inverse problem, which in this case is particularly challenging because one of the GPD variables (the average parton momentum fraction $x$) is completely integrated out and does not appear as a kinematic variable for the CFFs.

In Refs.~\cite{Bertone:2021yyz, Bertone:2021wib}, it has been shown that the inverse problem associated with inferring GPDs from DVCS data can yield an infinite set of solutions or ``possible GPDs'' at a fixed value of the scale of the reaction $Q^2 = -q^2$, which is minus the square of the four-momentum transfer from the incident lepton. These multiple solutions are characterized by what have been called shadow GPDs (SGPDs), which are infinite classes of functions that are orthogonal to the various hard-scattering coefficient functions and therefore do not contribute to the CFFs. The various classes of SGPDs are process-dependent and are also subject to the constraints associated with the formal properties of GPDs, such as polynominality, correct support, and sum rules. Nevertheless, SGPDs can still be very large and could render impossible the extraction of GPDs from DVCS data.

To adequately constrain SGPDs and fully infer GPDs from data will likely require a multifaceted approach, which in addition to DVCS will include data from other processes that are sensitive to GPDs, such as deeply virtual meson production (DVMP)~\cite{Favart:2015umi,Diehl:2003ny,Belitsky:2005qn}, double DVCS~\cite{Guidal:2002kt,Belitsky:2005qn}, and other processes directly sensitive to the $x$ dependence of GPDs~\cite{Duplancic:2018bum,Qiu:2022bpq,Qiu:2022pla,Siddikov:2022bku}. A complementary approach is to also exploit the correlation between $x$ and $Q^2$ that is provided by the QCD evolution of GPDs and hence SGPDs. This is made possible because the known classes of SGPDs at a particular $Q^2$ do evolve with scale, and in so doing are no longer orthogonal to the hard-scattering coefficients and thereby begin to contribute to the CFFs upon evolution. This implies that, in principle, it is likely possible to completely constrain the SGPDs with perfect CFF data over a finite $Q^2$ range.

In this work we study the impact that QCD evolution can have on constraining SGPDs when inferring GPDs from DVCS data that has kinematics similar to those at Jefferson Lab, COMPASS, and the EIC. This paper is organized as follows: Sec.~\ref{sec:formalism} reviews the formalism associated with CFFs, GPDs, and SGPDs. Sec.~\ref{sec:proxy} introduces a proxy model for the GPDs that is needed for our analysis. In \sref{sgpdev} we use simulated CFF data to study the impact that scale evolution can have on constraining the SGPDs. \sref{sgpdimpact} presents our findings regarding the impact SGPDs could have on determining the total angular momentum, pressure and shear force distributions, and tomography. Finally, we summarize our results and provide conclusions in \sref{conclusion}.

\section{CFFs, GPDs, AND SGPDs\label{sec:formalism}}
For spin-half targets such as the nucleon, there are four complex-valued CFFs ($\mathcal{H}$, $\mathcal{E}$, $\tilde{\mathcal{H}}$, $\tilde{\mathcal{E}}$) that enter the DVCS cross section at leading-twist, which are related to several quark and gluon leading-twist GPDs ($H^a$, $E^a$, $\tilde{H}^a$, $\tilde{E}^a$) by~\cite{Belitsky:2001ns}
\begin{align}
\label{e.gpd2cff}
\mathcal{F}(\xi,t,Q^2) 
&= \int_{-1}^1 \dd x \ \sum_a C^a(x,\xi,Q^2,\mu^2)\, F^a(x,\xi,t;\mu^2),
\end{align}
where $\mathcal{F} = \mathcal{H},~\mathcal{E}$ are associated with the leading-twist spin-independent GPDs $F = H^a,~E^a$, respectively, the sum is over all active parton flavors ($a = q,\,g$), and $C^a$ is a hard-scattering coefficient function. An analogous relation holds between the $\tilde{\mathcal{H}}$ and $\tilde{\mathcal{E}}$ CFFs and the leading-twist spin-dependent GPDs $\tilde{H}^a$ and $\tilde{E}^a$, where in this case different hard-scattering coefficient functions $(\tilde{C}^a)$ enter (see App.~\ref{ss.CFFs})~\cite{Belitsky:2001ns}. The GPDs are formally defined through matrix elements of quark and gluon operators at a light-like separation. For a spin-half target like the nucleon the leading twist unpolarized quark and gluon GPDs are defined by~\cite{Ji:1998pc,Diehl:2003ny}\footnote{We use the Ji convention~\cite{Ji:1998pc} for the gluon GPDs, which differs from the Diehl convention~\cite{Diehl:2003ny} by $F^g_{\rm Diehl} = 2\,x\,F^g_{\rm Ji}$.}
\begin{align}
\label{e.qGPD}
P\cdot n\int\frac{\dd \lambda}{2\pi}\ e^{ix P\cdot n\lambda}
\left< p'\lf|\bar{\psi}^q(-\tfrac12 \lambda n)\,\sh{n}\, \psi^q (\tfrac12 \lambda n)\rg|p\right>
&= 
\bar{u}(p')
\left[\ph{x\,}H^q(x,\xi,t;\mu^2)\,\sh{n} + 
\ph{x\,}E^q(x,\xi,t;\mu^2)\,\frac{i\sigma^{n \Delta}}{2M}
\right]u(p), \\
\label{e.gGPD}
n_\mu n_\nu \int\frac{\dd \lambda}{2\pi}\ e^{ix P\cdot n\lambda}
\big< p'\lf|G^{\mu\alpha}(-\tfrac12 \lambda n)\,{G_\alpha}^\nu(\tfrac12 \lambda n)\rg|p \big>
&= \bar{u}(p') 
\left[x\,H^g(x,\xi,t;\mu^2)\, \sh{n}
+ x\,E^g(x,\xi,t;\mu^2)\, \frac{i\sigma^{n \Delta}}{2M} 
\right]u(p),
\end{align}
where $n$ is a light-like vector ($n^2=0$) and analogous expressions hold for the polarized GPDs $\tilde{H}^a$ and $\tilde{E}^a$ (see  App.~\ref{app:gpds})~\cite{Diehl:2003ny,Freese:2020mcx}. For legibility we do not display the polarization dependence of the hadron states and spinors nor the color degrees of freedom in the field operators (which should be implicitly understood as containing gauge links and color sums). We use the notation $\sigma^{n \Delta} \equiv \sigma^{\mu\nu}n_\mu \Delta_\nu$, $M$ is the target mass, $P = \tfrac{1}{2}(p'+p)$ is the average of the initial and final target momenta, $\Delta = p' - p$ is the momentum transferred to the target, $t=\Delta^2$, $x = k\cdot n/P\cdot n$ is the average light cone momentum fraction of the active parton where $k$ is the average of the initial and final parton momenta, and $\xi=-\Delta \cdot n/(2P\cdot n)$ is a measure of the longitudinal momentum transfer to the target (skewness). The GPDs also depend on the renormalization scale $\mu^2$ in accordance with the renormalization group  equations~\cite{Ji:1996nm,Radyushkin:1997ki,Vinnikov:2006xw}. GPDs have support in the region $-1 \leqslant x \leqslant 1$ and $-1 \leqslant \xi \leqslant 1$, with the constraint that for a given $\xi$ the maximal value of $t \leqslant t_{\rm min} \leqslant 0$ for a physical process is $t_{\rm min} = -4\xi^2M^2/(1-\xi^2)$.

The GPDs exhibit several interesting properties, such as polynomiality~\cite{Diehl:2003ny}, which is a consequence of Lorentz covariance and implies the $x$-weighted moments of GPDs are even polynomials in $\xi$ (see App.~\ref{app:gpds} for more details). GPDs also formally satisfy numerous positivity conditions~\cite{Pobylitsa:2001nt,Pobylitsa:2002gw,Kirch:2005in}, with an example for the nucleon including
\begin{align}
\left|H^q(x,\xi,t;\mu^2) - \frac{\xi^2}{1-\xi^2}\,E^q(x,\xi,t;\mu^2)\right|^2
+ \left|\frac{\sqrt{t_{\rm min} - t}}{2M\sqrt{1-\xi^2}}\,E^q(x,\xi,t;\mu^2)\right|^2
&\leqslant \frac{q(x_{\rm in};\mu^2)\,q(x_{\rm out};\mu^2)}{1-\xi^2}, \hspace*{10mm}
\label{eq:pos}
\end{align}
where $q(x;\mu^2)$ are the familiar collinear PDFs, $x_{\rm in} = (x+\xi)/(1+\xi)$, $x_{\rm out} = (x-\xi)/(1-\xi)$, and this positivity constraint applies in the region $|x| > |\xi|$. In the forward limit ($\xi \to 0, t \to 0$) the $H^a$ and $\tilde{H}^a$ GPDs reduce to~\cite{Belitsky:2001ns}
\begin{align}
\label{eq:forward1}
H^q(x,0,0) &= q(x)\,\Theta(x) - \bar{q}(-x)\,\Theta(-x),  &
\tilde{H}^q(x,0,0) &= \Delta q(x)\,\Theta(x) + \Delta \bar{q}(-x)\,\Theta(-x), \\
\label{eq:forward2}
2\,H^g(x,0,0) &= g(x)\,\Theta(x) - g(-x)\,\Theta(-x), & 
2\,\tilde{H}^g(x,0,0) &= \Delta g(x)\,\Theta(x) + \Delta g(-x)\,\Theta(-x),
\end{align}
where we have dropped the $\mu^2$ dependence, the factor 2 for gluons avoids double counting because gluons are their own anti-particle, $\Theta$ is the Heaviside step function, and $q$, $\bar{q}$, and $g$ are the PDFs for the quarks, anti-quarks, and gluons, respectively. The first moments of quark GPDs are related to the quark contributions to the Dirac, Pauli, axial, and pseudoscalar form factors:
\begin{align}
\int_{-1}^1\dd{x}\,[H^q,E^q,\tilde{H}^q,\tilde{E}^q](x,\xi,t;\mu^2) &= 
[F^q_1(t),F^q_2(t),G^q_A(t),G_P^q(t)].
\label{eq:gpdsffs}
\end{align}
An important reason for the interest in GPDs, is that their second moments are related to the quark and gluon gravitational form factors. For the nucleon this implies~\cite{Polyakov:2018zvc}
\begin{align}
\int_{-1}^1\dd{x}\ x\,H^a(x,\xi,t;\mu^2) &= A^a(t;\mu^2) + \xi^2\,C^a(t;\mu^2), &
\int_{-1}^1\dd{x}\ x\,E^a(x,\xi,t;\mu^2) &= B^a(t;\mu^2) - \xi^2\,C^a(t;\mu^2), 
\label{eq:emtffs}
\end{align}
where the form factors are defined with respect to matrix elements of the energy-momentum tensor as
\begin{align}
\lf<p',\lambda'\lf|T^{\mu\nu}_a(x)\rg|p,\lambda\rg> &=
\bar{u}(p',\lambda')
\lf[A^a(t)\, \frac{\gamma^{\{\mu}P^{\nu\}}}{2}
+ B^a(t)\,\frac{P^{\{\mu}i\sigma^{\nu\}\Delta}}{4\,M}
+ C^a(t)\,\frac{\Delta^\mu \Delta^\nu - \Delta^2\,g^{\mu\nu}}{4\,M}
+ M\,\bar{C}^a(t)\,g^{\mu\nu}
\rg]u(p,\lambda).
\label{eq:emtcurrent}
\end{align}
We have introduced the notation $a^{\{\mu}b^{\nu\}} = a^\mu b^\nu + a^\nu b^\mu$. The total quark and gluon angular moment is then given by the Ji sum rule~\cite{Ji:1996ek} as $J^a(\mu^2) = \tfrac{1}{2}[A^a(0,\mu^2) + B^a(0,\mu^2)]$ and $J = \tfrac{1}{2} = \sum_a\,J^a(\mu^2)$. The $C^a$ form factors are related to internal stresses within the nucleon~\cite{Polyakov:2002yz,Polyakov:2018zvc,Lorce:2018egm,Freese:2021czn,Dutrieux:2021nlz,Lorce:2021xku}, with $D = C(0) = \sum_a\,C^a(0;\mu^2)$ known as the nucleon $D$-term, and $\sum_a\bar{C}^a(t,\mu^2) = 0$.

It is made clear by Eq.~\eqref{e.gpd2cff} that inferring GPDs from DVCS data involves solving several inverse problems, subject to some or all of the constraints given by Eqs.~\eqref{eq:pos}--\eqref{eq:emtffs}~\cite{Freund:1999xf,Kumericki:2016ehc,Guo:2022upw,Guo:2023ahv}. The first step in this procedure is to obtain the CFFs from DVCS data, which at leading twist is a closed problem~\cite{Belitsky:2001ns} and several leading-twist CFF extractions have been reported in the literature~\cite{Moutarde:2019tqa,Cuic:2020iwt,Kriesten:2020apm,Grigsby:2020auv,Shiells:2021xqo,Almaeen:2022imx}. The challenge of extracting GPDs from the CFFs lies in the fact that the $x$ dependence of the GPDs is completely integrated out and does not appear in the CFFs. Nevertheless, GPD extractions are beginning to become available; see, for instance, Ref.~\cite{Guo:2023ahv}.

In Refs.~\cite{Bertone:2021yyz,Bertone:2021wib}, it has recently been shown that, at a fixed value of $\mu^2 = Q^2$ in Eq.~\eqref{e.gpd2cff} and subject to the forward-limit and form factor constraints of Eqs.~\eqref{eq:forward1}--\eqref{eq:gpdsffs}, it is not possible to uniquely extract the $H^a$ GPDs from DVCS data. This is because there are multiple solutions to the inverse problem of Eq.~\eqref{e.gpd2cff}, that each satisfy the forward limit and electromagnetic form factor constraints. The analysis in Refs.~\cite{Bertone:2021yyz,Bertone:2021wib} was conducted at leading order (LO) and next-to-leading order (NLO) in the hard scattering coefficients and also straightforwardly applies to the $E^a$, $\tilde{H}^a$, and $\tilde{E}^a$ GPDs. We label these multiple solutions as $F^a_F(x,\xi,t;\mu^2)$, with the subscript $F$ indicating a ``false'' GPD which, nevertheless, satisfies all requirements of a GPD, e.g., polynominality, forward limit, etc. Then, without loss of generality, these false GPDs can be expressed as
\begin{align}
F^a_F(x,\xi,t;\mu^2) &= F^a_T(x,\xi,t;\mu^2) + F^a_S(x,\xi,t;\mu^2),
\end{align}
where the unique GPD associated with its matrix element definition, such as those given in Eqs.~\eqref{e.qGPD}--\eqref{e.gGPD}, is labeled by $F^a_T(x,\xi,t;\mu^2)$ with the subscript $T$ indicating the ``true'' GPD, and $F^a_S(x,\xi,t;\mu^2)$ are the so-called SGPDs that characterize the difference between the true and false GPDs. The SGPDs, by construction, do not contribute to any experimentally accessible observable, which implies, e.g., that $H^a_S$ and $\tilde{H}^a_S$ must vanish in the forward limit in accordance with Eqs.~\eqref{eq:forward1}--\eqref{eq:forward2} and the first moments of $H^a_S,E^a_S,\tilde{H}^a_S$, and $\tilde{E}^a_S$ must vanish so that Eqs.~\eqref{eq:gpdsffs} remain valid. In addition, SGPDs do not contribute to the CFFs, which implies
\begin{align}
\sum_a C^a(x,\xi,Q^2,\mu^2)\, F^a_S(x,\xi,t;\mu^2) = 0,
\label{eq:shadow}
\end{align}
where the specific hard-scattering coefficient function $C^a$ depends on the process and type of GPD. Therefore, SGPDs make it impossible to distinguish the ``true'' GPDs, $F^a_T$, from the sum of the ``true'' GPDs and SGPDs, $F^a_T + F^a_S$, using only DVCS data at a fixed scale $Q^2$. As shown in Refs.~\cite{Bertone:2021yyz, Bertone:2021wib}, the shape and size of $F^a_F = F^a_T + F^a_S$ may differ significantly from $F^a_T$. 

An important caveat is that the known classes of SGPDs do not satisfy Eq.~\eqref{eq:shadow} after QCD evolution, and therefore begin to contribute to CFFs upon evolution to a different scale. Therefore, a SGPD at an initial scale $\mu^2_0$ is no longer a SGPD when evolved to a new scale $\mu^2$. It is therefore plausible that QCD evolution of the GPDs has the potential to help constrain SGPDs and make viable the extraction of GPDs from DVCS data with a sufficiently large $Q^2$ range. We note that using QCD evolution to help infer GPDs from data was first suggested by Freund~\cite{Freund:1999xf}. 

The original SGPD analysis presented in Refs.~\cite{Bertone:2021yyz, Bertone:2021wib} suggested that the impact of evolution should be very small. We extend this analysis by explicitly solving the evolution equations for the GPDs, and study whether QCD evolution may be capable of providing constraints on SGPDs and render the extraction of GPDs from DVCS data practical.

\section{GPD PROXY MODEL AND A CLASS OF SGPDS }
\label{sec:proxy}
To investigate the ability of QCD evolution to constrain the SGPDs, and thereby extract information on the true GPDs from DVCS data, it is necessary to adopt a proxy model for these true GPDs. We chose the widely-used phenomenological model from Vanderhaeghen, Guichon, and  Guidal (VGG)~\cite{Vanderhaeghen:1998uc, Vanderhaeghen:1999xj, Goeke:2001tz, Guidal:2004nd}, however, the outcome of our analysis does not directly depend on this choice as the SGPDs are independent of the proxy model. To ensure the polynomiality property of the GPDs it is common to work at the level of double distributions~\cite{Radyushkin:1997ki}. Our analysis will focus on the  leading-twist spin-independent nucleon GPDs, which in terms of double distributions are given by~\cite{Diehl:2003ny}
\begin{align}
\label{eq:HDD}
H^a(x,\xi,t;\mu^2) &= \int\dd\beta\dd\alpha\  \delta(x-\beta-\xi\alpha) \big[H_{DD}^a(\beta,\alpha,t;\mu^2) + \delta(\beta)\,\xi\,D^a(\alpha,t;\mu^2)
\,\Theta\left(\left|\xi\right| - \left|x\right|\right)\big], \\  
\label{eq:EDD}
E^a(x,\xi,t;\mu^2) &= \int\dd\beta\dd\alpha\  \delta(x-\beta-\xi\alpha) \big[E_{DD}^a(\beta,\alpha,t;\mu^2) - \delta(\beta)\,\xi\,D^a(\alpha,t;\mu^2)
\,\Theta\left(\left|\xi\right| - \left|x\right|\right)\big],
\end{align}
where $|\alpha|+|\beta| \leqslant 1$, the functions $H_{DD}^a$ and $E_{DD}^a$ are the double distributions which must be even in $\alpha$ to preserve polynomiality, and $D^a(\alpha,t;\mu^2)$ is called the D-term \cite{Polyakov:1999gs}. The function $D^a(\alpha,t;\mu^2)$ shares the same ``D-term'' name as the value of the gravitational form factor $C(t;\mu^2) = \sum_a C^a(t;\mu^2)$ at $t=0$, and they are related by $C^a(t;\mu^2) = \int_{-1}^1\dd \alpha\, \alpha\,D^a(\alpha,t;\mu^2)$.  

In the VGG model, and other models such as those from Goloskokov and Kroll~\cite{Goloskokov:2005sd,Goloskokov:2007nt,Goloskokov:2009ia}, the double distributions are parameterized at the input scale as
\begin{align}
\label{e.ddH} 
H_{DD}^q(\beta,\alpha,t;\mu_0^2) &= h(\beta,\alpha,t)\, 
  \big[~q(\beta;\mu_0^2)\Theta(\beta) - ~\bar{q}(-\beta;\mu_0^2)\Theta(-\beta) \big], &
H_{DD}^g(\beta,\alpha,t;\mu_0^2) &= \frac{1}{2}\,h(\beta,\alpha,t)\,
{\rm sgn}(\beta)\,g(|\beta|;\mu_0^2), \\
\label{e.dds} 
E_{DD}^q(\beta,\alpha,t;\mu_0^2) &= h(\beta,\alpha,t)\, 
  \big[e^{q}(\beta;\mu_0^2)\Theta(\beta) - e^{\bar{q}}(-\beta;\mu_0^2)\Theta(-\beta)\big], &
E_{DD}^g(\beta,\alpha,t;\mu_0^2) &= \frac{1}{2}\,h(\beta,\alpha,t)\, 
{\rm sgn}(\beta)\,e^g(|\beta|;\mu_0^2),
\end{align}
where $h(\beta,\alpha,t)$ is often called the profile function for which we take a standard form~\cite{Goeke:2001tz}:
\begin{equation} h(\beta,\alpha,t)=\frac{\Gamma(2b+2)}{2^{2b+1}\Gamma^2(b+1)}
\frac{\lf[(1-|\beta|)^2-\alpha^2\rg]^b}{(1-|\beta|)^{2b+1}}\ |\beta|^{-\omega(1-|\beta|)t}.
\label{e.prof}
\end{equation}
For the $q$, $\bar{q}$, and $g$ PDFs we take results from the JAM20-SIDIS global analysis~\cite{Moffat:2021dji} and in the profile function take the standard VGG parameters of $b=1$ and $\omega = 1.105$~\cite{Goeke:2001tz}, which provide a good fit to the nucleon's electromagnetic form factors. The forward limits of the $E^a$ GPDs do not correspond to any accessible collinear PDF, nevertheless, we use a common parametrization introduced in Ref.~\cite{Guidal:2004nd} for $E^q_{DD}$ where:
\begin{align}
e^u(x;\mu^2_0) &= \frac{\kappa_u}{N_u}\, (1-x)^{\eta_u}\, u_v(x;\mu^2_0), &
e^d(x;\mu^2_0) &= \frac{\kappa_d}{N_d}\, (1-x)^{\eta_d}\, d_v(x;\mu^2_0).
\end{align}
The empirical quark anomalous magnetic moments have the values $\kappa_u=1.673$ and $\kappa_d=-2.033$. For the $u_v$ and $d_v$ PDFs we again use those from Ref.~\cite{Moffat:2021dji}, and the normalization constants $N_q$ are fixed by requiring that the first moment of the associated $e^q$ functions give $\kappa_q$. This is necessary to reproduce the Pauli form factors of the nucleon at $t=0$. Following the VGG model we use $\eta_u=1.713$ and $\eta_d=0.566$, and set the remaining $e^a(x;\mu^2)$ functions to zero.\footnote{The functions $e^a$ defined in Eq.~\eqref{e.dds} should not be confused with the twist-3 PDFs that use the same symbol~\cite{Jaffe:1991kp}.} In our proxy model for the true GPDs we will omit the D-term because the known classes of SGPDs derived in Ref.~\cite{Bertone:2021yyz} have no D-term. This choice does not impact the results of our SGPD analysis, and is also motivated by the observable CFFs which can directly constrain the D-term via dispersion relations~\cite{Anikin:2007yh,Diehl:2007jb,Bertone:2021yyz}. See App.~\ref{ss.CFFs} for further details.

\begin{figure}
\includegraphics[width=\columnwidth]{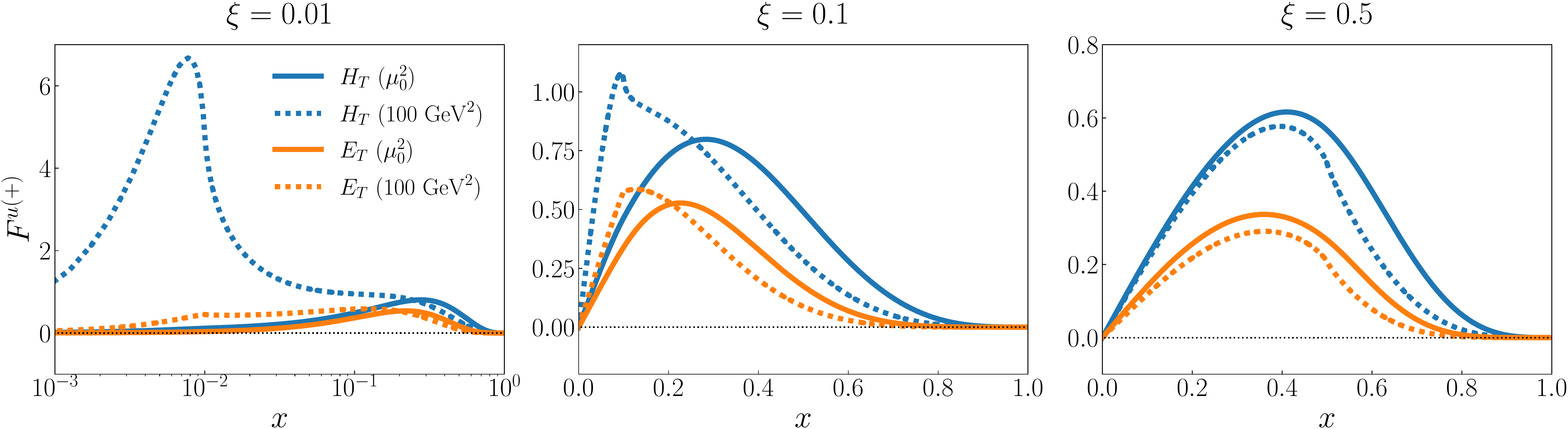}
\caption{Our VGG model results for the $H^{u(+)}$ (blue) and $E^{u(+)}$ (orange) GPDs, with $t=-1.2\,$GeV$^2$, and from left to right the skewness values are $\xi=0.01,\,0.1,\,0.5$. The input scale results (solid lines) correspond to $\mu_0 =1.27\,$GeV, taken from the JAM-SIDIS analysis~\cite{Moffat:2021dji}, and the dotted lines are these results evolved at LO to the scale $\mu^2=100$~GeV$^2$.}
\label{f.model}
\end{figure}

In this analysis we focus on the $\mathcal{H}$ and $\mathcal{E}$ CFFs and associated spin-independent GPDs and SGPDs. For these CFFs the hard-scattering coefficient functions [see Eq.~\eqref{e.gpd2cff}] for the associated GPDs are odd under $x \to -x$, and therefore only the singlet combination $F^{a(+)} = F^a(x,\xi,t) - F^a(-x,\xi,t)$ contributes to the CFFs. In \fref{model}, we present our VGG model results for the $u$ quark singlet $H^{u(+)}$ and $E^{u(+)}$ GPDs, where we take $t=-1.2\,$GeV$^2$ in Eq.~\eqref{e.prof} which accommodates the illustrated skewness values of $\xi=0.01,\,0.1,\,0.5$. Similar results are easily obtained for the other $H^{a(+)}$ and $E^{a(+)}$ GPDs. Since the GPD singlet combination is anti-symmetric in $x$ we just present proxy model GPDs for $0 \leqslant x \leqslant 1$. In Fig.~\ref{f.model} the GPDs are shown at the input scale of $\mu_0=1.27\,$GeV associated with the JAM20-SIDIS PDFs and after LO evolution to $\mu^2=100$~GeV$^2$. Details on the implementation of the GPD evolution are given in App.~\ref{s.evolution}.

To construct our SGPDs we will generally follow the same method as Ref.~\cite{Bertone:2021yyz}. Firstly, the class of SGPDs that we consider will be assumed separable such that
\begin{align}
F^a_S(x,\xi,t;\mu^2) \to F^a_S(x,\xi;\mu^2)\ f_S^a(t).
\label{eq:gpdsep}
\end{align}
To construct a general class of singlet SGPDs, we begin with a double distribution parametrization of the polynomial form~\cite{Bertone:2021yyz}
\begin{equation}
\label{e.sdd}
F_{S,DD}^{a(+)}(\beta,\alpha,t;\mu^2_0) = \sum_{m=0,\,2,\ldots;~n=1,\,3,\ldots}^{m+n \leqslant N}
c_{mn}^{a(+)}(\mu_0^2)\ \alpha^m\beta^n\ f_S^a(t),
\end{equation}
which holds at an initial scale $\mu_0^2$. Polynomiality requires $m$ to be even, while $n$ is taken as odd to give SGPDs that are anti-symmetric in $x$ so that they contribute to the singlet GPD combination. This choice also guarantees that the SGPDs do not contribute to the form factors via Eqs.~\eqref{eq:gpdsffs}. The $t$ dependence of the shadow double distributions, and therefore the SGPDs, is represented by $f_S^a(t)$ and for this analysis is left undefined. The exception is in the forward limit where we will consider two cases, corresponding to $f_S^a(0) \neq 0$ and $f_S^a(0) = 0$. The positive odd integer $N$ gives the order of the double distribution that has $(N+1)(N+3)/8$ unknown coefficients $c_{mn}$. The SGPDs associated with $F_{S,DD}^{a(+)}(\beta,\alpha,t;\mu^2_0)$ are then obtained by inserting \eref{sdd} into either Eq.~\eqref{eq:HDD} or \eqref{eq:EDD} to obtain $F^{a(+)}_S(x,\xi,t;\mu^2_0)$.

The key relation for the SGPDs is Eq.~\eqref{eq:shadow}, which is a set of homogeneous equations that place constraints on the coefficients $c_{mn}$ in Eq.~\eqref{e.sdd}, up to an overall normalization. For the known classes of SGPDs, Eq.~\eqref{eq:shadow} can only be implemented at a fixed scale $\mu^2_0$ and is no longer satisfied after evolution of the GPDs. Additional constraints on the SGPDs for $H^a$ and $\tilde{H}^a$ are given by the forward limit ($\xi \to 0,\,t \to 0$) results of Eqs.~\eqref{eq:forward1} and \eqref{eq:forward2}. This means the SGPDs for $H^a$ and $\tilde{H}^a$ must vanish in the forward limit, which can be implemented in two ways: 1) Via constraints on the $c_{mn}$ coefficients which means these SGPDs vanish when $\xi \to 0$; or 2) Choose the $t$ dependence of these SGPDs such that $f_S^a(0) =0$. We will investigate both of these scenarios, which we refer to as type A and type B SGPDs, respectively. As we will see, the type B SGPDs are less constrained which has implications for inferring GPDs from DVCS data. To satisfy the SGPD constraints a minimum order $N$ in Eq.~\eqref{e.sdd} is required, where in general there are more coefficients $c_{mn}$ than constraints making this an underdetermined problem. Therefore, for a sufficiently large $N$ there are in general an infinite number of SGPDs. To generate examples of SGPDs of order $N$, we need to reduce the number of unknown coefficients $c_{mn}$ to equal the number of constraint equations. This is done by assigning random values to the surplus coefficients, which are themselves randomly selected from the full set. The constraint equations are then solved to obtain the remaining coefficients.

\begin{figure}
\includegraphics[width=\columnwidth]{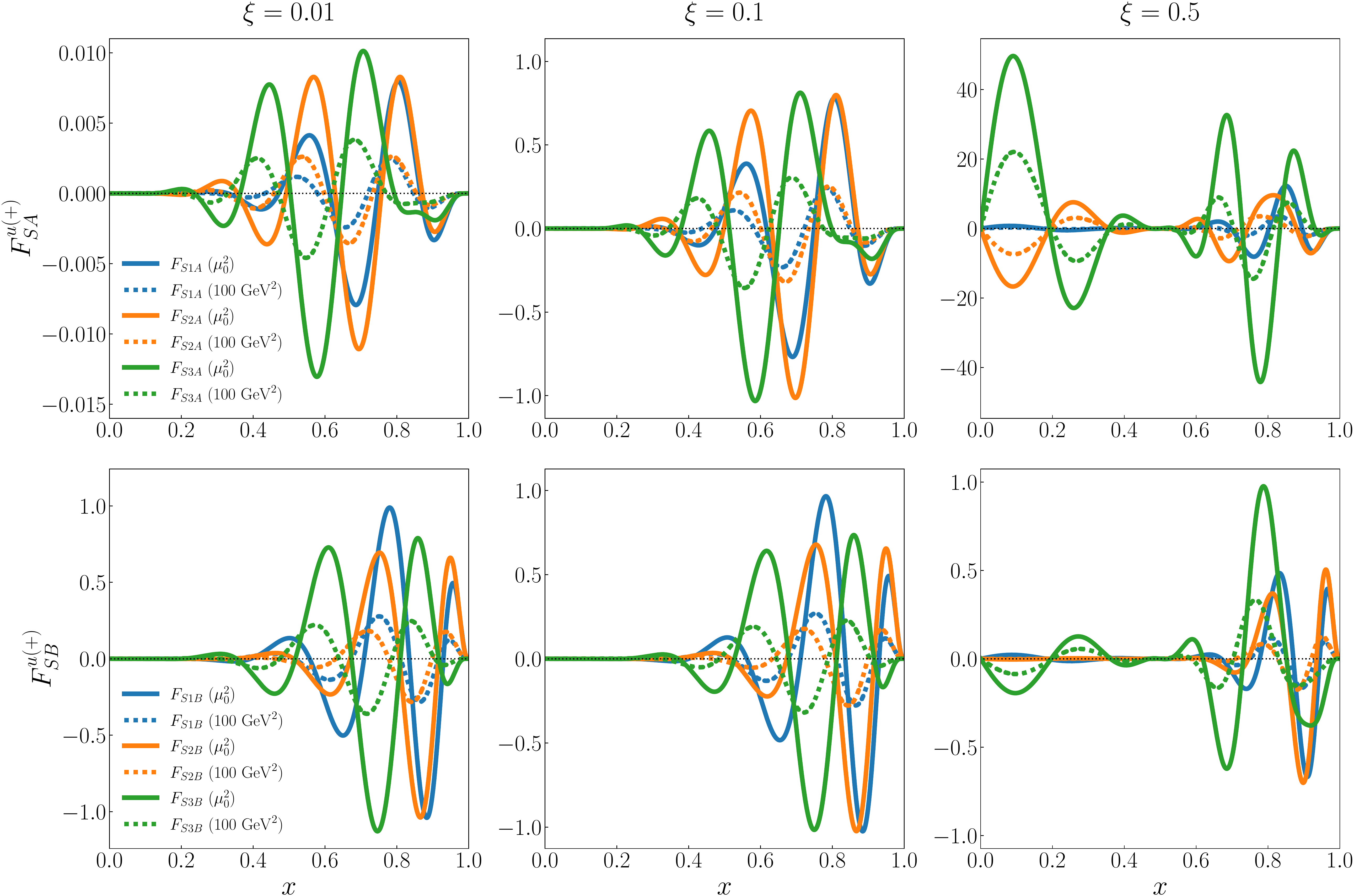}
\caption{Results for NLO SGPDs in the spin-independent quark sector. The top row shows three examples of type A SGPDs where the forward limit is maintained via constraints on the $c_{mn}$'s so that the SGPDs vanish as $\xi \to 0$. In the second row, three type B SGPDs are illustrated where any forward-limit constraint is imposed via the unspecified $t$ dependence. In each case, SGPDs with skewness $\xi=0.01,\,0.1,0.5$ are shown in the panels from left to right. The SGPDs are defined at the input scale $\mu_0 = 1.27\,$GeV (solid lines). These results are evolved at LO to $\mu^2=100$~GeV$^2$ (dotted lines), where these evolved results no longer satisfy the requirements for a SGPD.}
\label{f.NLOHS}
\end{figure}

Fig.~\ref{f.NLOHS} presents example SGPDs $F^{q(+)}_S(x,\xi;\mu^2)$ [see Eq.~\eqref{eq:gpdsep}] appropriate for the $H^{q(+)}$ and $E^{q(+)}$ GPDs, where in Eq.~\eqref{e.sdd} we have taken $N=27$, which is the same order as the example SGPDs discussed in Ref.~\cite{Bertone:2021yyz}. The top row in Fig.~\ref{f.NLOHS} gives results for type A SGPDs. (Note that in Ref.~\cite{Bertone:2021yyz} only type A SGPDs were explored.) In the second row of Fig.~\ref{f.NLOHS} this constraint on the $c_{mn}$'s is relaxed, and we illustrate results for type B SGPDs. These type B SGPDs are also appropriate for the GPD $E^q$, as in this case there is no forward-limit constraint. For each type of SGPD we give results for three values of skewness, $\xi=0.01,\,0.1,\,0.5$, and the input scale is $\mu_0 = 1.27\,$GeV. We also show results where these SGPDs are evolved at LO to $\mu^2 = 100\,$GeV$^2$, where importantly, these results no longer satisfy Eq.~\eqref{eq:shadow} and are therefore not SGPDs and can in principle be constrained by DVCS data.

Since Eq.~\eqref{eq:shadow} is homogeneous, the SGPDs can have an arbitrary normalization, so for the results in Fig.~\ref{f.NLOHS} and in the following analysis we fixed the normalization of the SGPDs such that ${\rm max}[|F^{q(+)}_S(x,\xi=0.1;\mu_0^2)|] \simeq 1$. From Fig.~\ref{f.NLOHS}, it is clear that the type A SGPDs in the top row get very small as $\xi \to 0$, because of how the forward limit is imposed, whereas for large $\xi$ these SGPDs tend to grow rapidly. Results for the type B SGPDs in the second row, tend to be of similar size independent of $\xi$. Analogous results for gluon and spin-dependent SGPDs are easily obtained using procedures similar to those described above.

\section{SIMULATED CFF DATA, SHADOW GPDS, AND QCD EVOLUTION}
\label{s.sgpdev}
\begin{figure}
\includegraphics[width=\columnwidth]{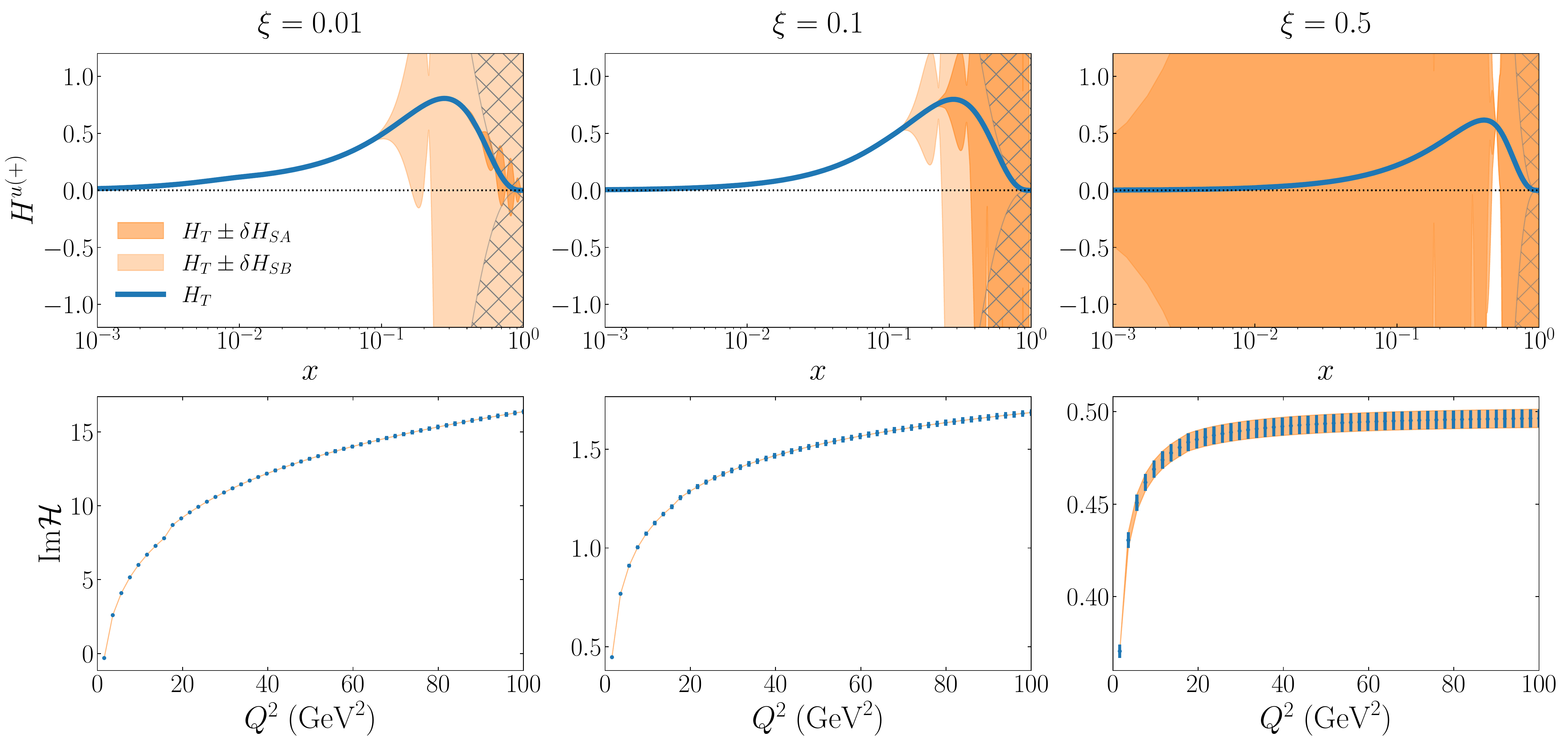}
\caption{{\it Top row}: Results for the true GPD $H_T^{u(+)}$ (blue lines) and the range of $H_T^{u(+)} \pm \delta{H}_S$ (orange bands) at the input scale for three different values of $\xi$.  The darker orange bands show the results for the three type A SGPDs in \fref{NLOHS}, while the lighter orange bands show the results for the type B SGPDs in \fref{NLOHS}.  The gray hashed region shows the area that would be excluded by the positivity constraint for $H^q$ given in Eq.~\eqref{eq:Hpos2}. 
{\it Bottom row}: Simulated data (blue points) for the CFF ${\rm Im}{\mathcal H}$ obtained using the VGG model GPDs with error bars showing 1\% uncertainty, as a function of $Q^2$, for fixed $t$, and three values of $\xi$. The dark orange bands illustrate the 10000 sample CFFs associated with type A SGPDs.}
\label{f.NLOsgpds}
\end{figure}

\begin{figure}
\includegraphics[width=\columnwidth]{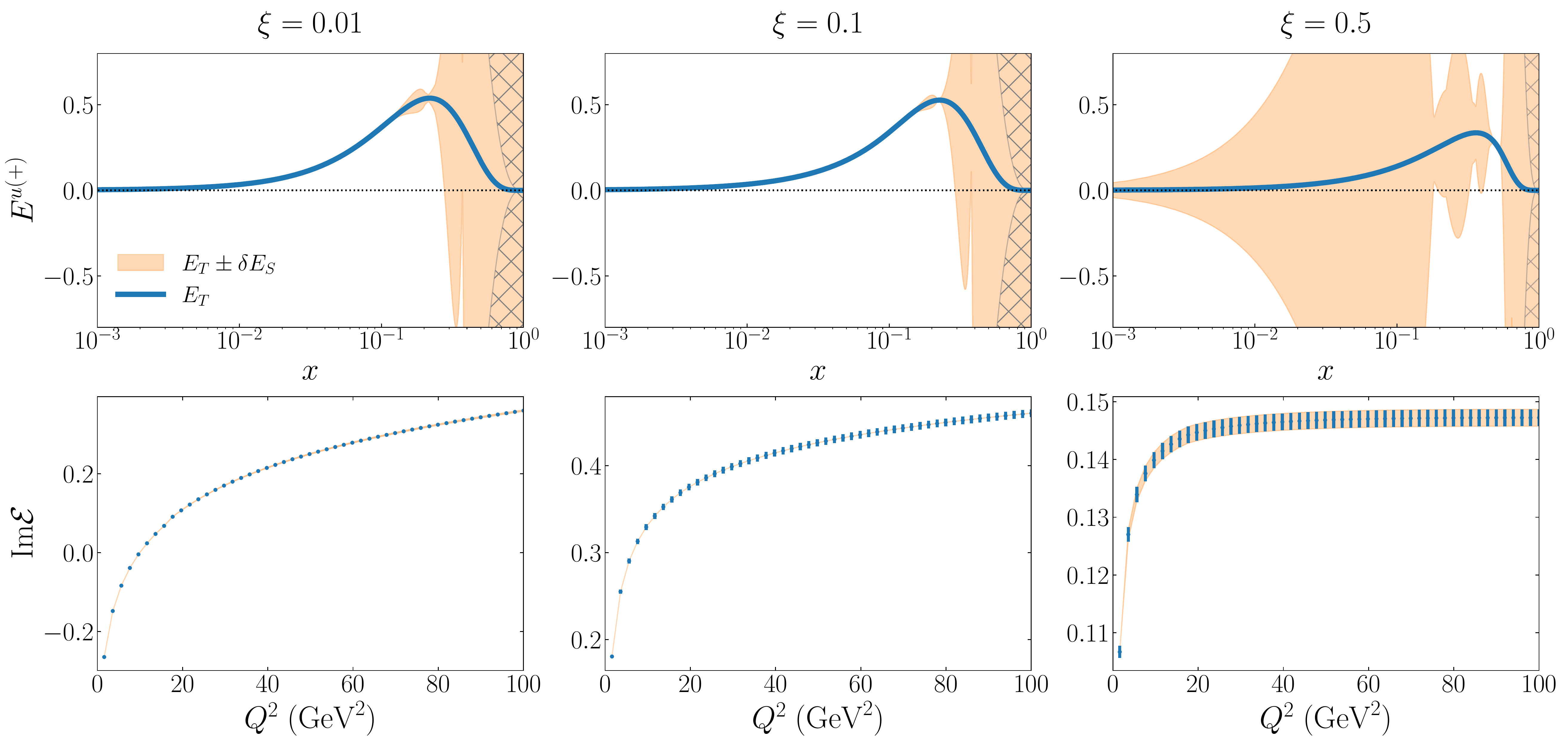}
\caption{{\it Top row}: Results for the true GPD $E_T^{u(+)}$ (blue lines) and the range of $E_T^{u(+)} \pm \delta{E}_S$ (orange bands) at the input scale for three different values of $\xi$.  The orange bands show the results for the three type B SGPDs in \fref{NLOHS}, appropriate for $E^q$ GPDs because this is no forward limit constrain. The gray hashed region shows the area that would be excluded by the positivity constraint for $E^q$ given in Eq.~\eqref{eq:Hpos2}. 
{\it Bottom row}: Simulated data (blue points) for the CFF ${\rm Im}{\mathcal E}$ obtained using the VGG model GPDs with error bars showing 1\% uncertainty, as a function of $Q^2$, for fixed $t$, and three values of $\xi$. The light orange bands illustrate the 10000 sample CFFs associated with type B SGPDs.}
\label{f.NLOEsgpds}
\end{figure}

To replicate the inference of GPDs from DVCS data, we create simulated CFF data by taking the VGG model of Sec.~\ref{sec:proxy} as a set of ``true'' GPDs. Then setting $\mu^2 = Q^2$ we use QCD evolution of these GPDs, together with Eq.~\eqref{e.gpd2cff} and the hard-scattering coefficients given in App.~\ref{ss.CFFs}, to construct simulated results for ${\rm Im}{\mathcal H}(\xi,t,Q^2)$ and ${\rm Im}{\mathcal E}(\xi,t,Q^2)$. Results for these CFFs, which include all active quark flavors and the gluons, are illustrated in the bottom rows of Figs.~\ref{f.NLOsgpds} and \ref{f.NLOEsgpds}. In particular, we have generated 50 data points for $Q^2$ values between the input scale of $Q_0^2 = (1.27$ GeV)$^2$ and $Q^2 = 100\,$GeV$^2$, and assigned a 1\% error to this simulated CFF data. These results were obtained by taking $t=-1.2\,$GeV$^2$ in the VGG model, which is an arbitrary choice and does not impact the conclusions drawn from our analysis but does accommodate the skewness values $\xi = 0.01,\,0.1,\,0.5$.

The goal of this analysis is to assess the impact that QCD evolution can have on constraining the contribution of the SGPDs, given simulated data for the imaginary parts of the CFFs $\mathcal{H}$ and  $\mathcal{E}$ over a wide range of kinematics. To implement this we will consider SGPDs in the $u$-quark singlet sector for the $H^{u(+)}$ and $E^{u(+)}$ GPDs. In this case the false GPDs are given by
\begin{align}
  \label{e.GPDHpar}
H^{u(+)}_F(x,\xi,t;\mu^2,\lambda_i) &= H^{u(+)}_T(x,\xi,t;\mu^2)
+ \lambda_1\, F^{q(+)}_{S1}(x,\xi;\mu^2)
+ \lambda_2\, F^{q(+)}_{S2}(x,\xi;\mu^2)
+ \lambda_3\, F^{q(+)}_{S3}(x,\xi;\mu^2), \\
  \label{e.GPDEpar}
E^{u(+)}_F(x,\xi,t;\mu^2,\lambda_i) &= E^{u(+)}_T(x,\xi,t;\mu^2)\,
+ \lambda_1\, F^{q(+)}_{S1}(x,\xi;\mu^2)
+ \lambda_2\, F^{q(+)}_{S2}(x,\xi;\mu^2)
+ \lambda_3\, F^{q(+)}_{S3}(x,\xi;\mu^2),
\end{align}
where $H^{u(+)}_T(x,\xi,t;\mu^2)$ and $E^{u(+)}_T(x,\xi,t;\mu^2)$ are the true $u$-quark GPDs from the VGG model that is discussed in Sec.~\ref{sec:proxy}, the $\lambda_{1,2,3}$ are arbitrary scaling parameters, and 
the $F^{q(+)}_{S}$ are the three example SGPDs illustrated in Fig.~\ref{f.NLOHS} which can be either type A or type B. We do not specify any $t$ dependence for the SGPDs but for $t < 0$ are basically assuming $f^a_S(t) = 1$. However, any explicit $t$ dependence could be absorbed into the $\lambda_{1,2,3}$, so our analysis can be considered general for any $t < 0$. For $H^{u(+)}_F$ we will study both type A and type B shadow GPDs, whereas for $E^{u(+)}_F$ we just consider type B as there is no forward-limit constraint for this GPD.

The SGPDs in Eqs.~\eqref{e.GPDHpar} and \eqref{e.GPDEpar} give zero contribution to CFFs at the input scale of $Q_0^2 = (1.27$ GeV)$^2$, however, upon evolution to a new scale $Q^2$ they do contribute to the CFFs. Therefore, having data over a finite $Q^2$ range makes it possible to constrain the SGPDs by putting limits on the values of  $\lambda_i$, and with perfect CFF data over a large kinematic range it would in principle be possible to completely constrain the SGPDs. We have assumed 1\% error on our simulated CFF data and we wish to explore the range of SGPDs ($\lambda_i$'s) that cannot be constrained by this data. To achieve this we generate replica SGPDs by randomly generating the three scaling parameters ($\lambda_{1,2,3}$) in the range $-10^4 \leqslant \lambda_i \leqslant 10^4$, and use these in Eqs.~\eqref{e.GPDHpar} and \eqref{e.GPDEpar} to generate replica false GPDs. These false GPDs are then combined with the true GPDs for the other flavors and the gluons to create sample CFFs for ${\rm Im}{\mathcal H}(\xi,t,Q^2)$ and ${\rm Im}{\mathcal E}(\xi,t,Q^2)$. If these sample CFFs are within the 1\% errors of the simulated CFF data, across all considered kinematics, then the associated replica cannot be ruled out by the data. We then repeat this process until we have 10000 sample SGPDs that cannot be ruled out by the CFF data, and from these SGPDs we define the quantity
\begin{align}
\delta{F}_S(x,\xi;\mu^2)
= {\rm max}\lf[\lf|\lambda_1\,F^{u(+)}_{S1}(x,\xi;\mu^2)
+ \lambda_2\,F^{u(+)}_{S2}(x,\xi;\mu^2)
+ \lambda_3\,F^{u(+)}_{S3}(x,\xi;\mu^2)\rg|\rg],
\label{eq:error}
\end{align}
which indicates how well the simulated CFF data can constrain the SGPDs, and therefore provides a measure for how accurately the true GPDs can be inferred from CFF data with 1\% uncertainties.

Results for the inferred $H^{u(+)}(x,\xi,t;\mu_0^2)$ GPD are illustrated in the top row of Fig.~\ref{f.NLOsgpds}, for $\xi=0.01,\,0.1,\,0.5$. The solid line illustrates the true GPD, $H^{u(+)}_T(x,\xi,t;\mu_0^2)$, and the shaded areas represent $H^{u(+)}_T \pm \delta H_S$. For $H^{u(+)}$ we consider both type A and type B SGPDs, where the dark orange shaded regions represent the constraints on the type A SGPDs which vanish in the limit $\xi \to 0$, while the light orange regions correspond to constraints on the type B SGPDs from the simulated CFF data. Analogous results for the true GPD, $E^{u(+)}_T(x,\xi,t;\mu_0^2)$, and results for $E^{u(+)}_T \pm \delta E_S$ are presented in the top row of Fig.~\ref{f.NLOEsgpds}. However, here only type B SGPDs are shown for $E^{u(+)}$ as this GPD has no forward limit contraint. For the $H^{u(+)}$ GPD, we see that at $\xi = 0.5$ both type A and type B SGPDs are only constrained at $x \simeq \xi$ and elsewhere the uncertainties are large. As skewness is decreased to $\xi = 0.1$ both type A and B SGPDs become well constrained for $x \lesssim 0.1$, while for larger $x$ the uncertainities remain large. Finally, for $\xi = 0.01$ type A become well constrained over the entire $x$ range, but type B SGPDs are still largely unconstrained for $x \gtrsim 0.2$. The constraints on type B SGPDs for $E^{u(+)}$ as a function of $x$ and $\xi$ are similar to those for type B SGPDs for the GPD $H^{u(+)}$. This is because the SGPDs are largely independent of the underlying proxy model.

To better understand how the simulated CFF data provides constraints on the SGPDs, in the bottom rows of Figs.~\ref{f.NLOsgpds} and \ref{f.NLOEsgpds} we illustrate as an orange band the 10000 sample CFFs that contain SGPDs that cannot be constrained by the simulated data. In Fig.~\ref{f.NLOsgpds} we do this for type A SGPDs and in Fig.~\ref{f.NLOEsgpds} for type B GPDs. For both ${\rm Im}{\mathcal H}(\xi,t,Q^2)$ and ${\rm Im}{\mathcal E}(\xi,t,Q^2)$ we notice that the orange band in the rightmost panel ($\xi=0.5$) is as wide as the uncertainty of the simulated CFF data, however, for the panels corresponding to $\xi=0.01,\,0.1$ the orange bands are generally narrower. This indicates that the simulated data at the largest skewness value is playing an important role in constraining the SGPDs for the smaller skewness values $\xi=0.01,\,0.1$. This is particularly the case for the type A SGPDs in Fig.~\ref{f.NLOsgpds}, which suggests that type A SGPDs can be well constrained at lower values of $\xi$ by evolution provided data is available at higher skewness values.

\begin{figure}
\includegraphics[width=0.8\columnwidth]{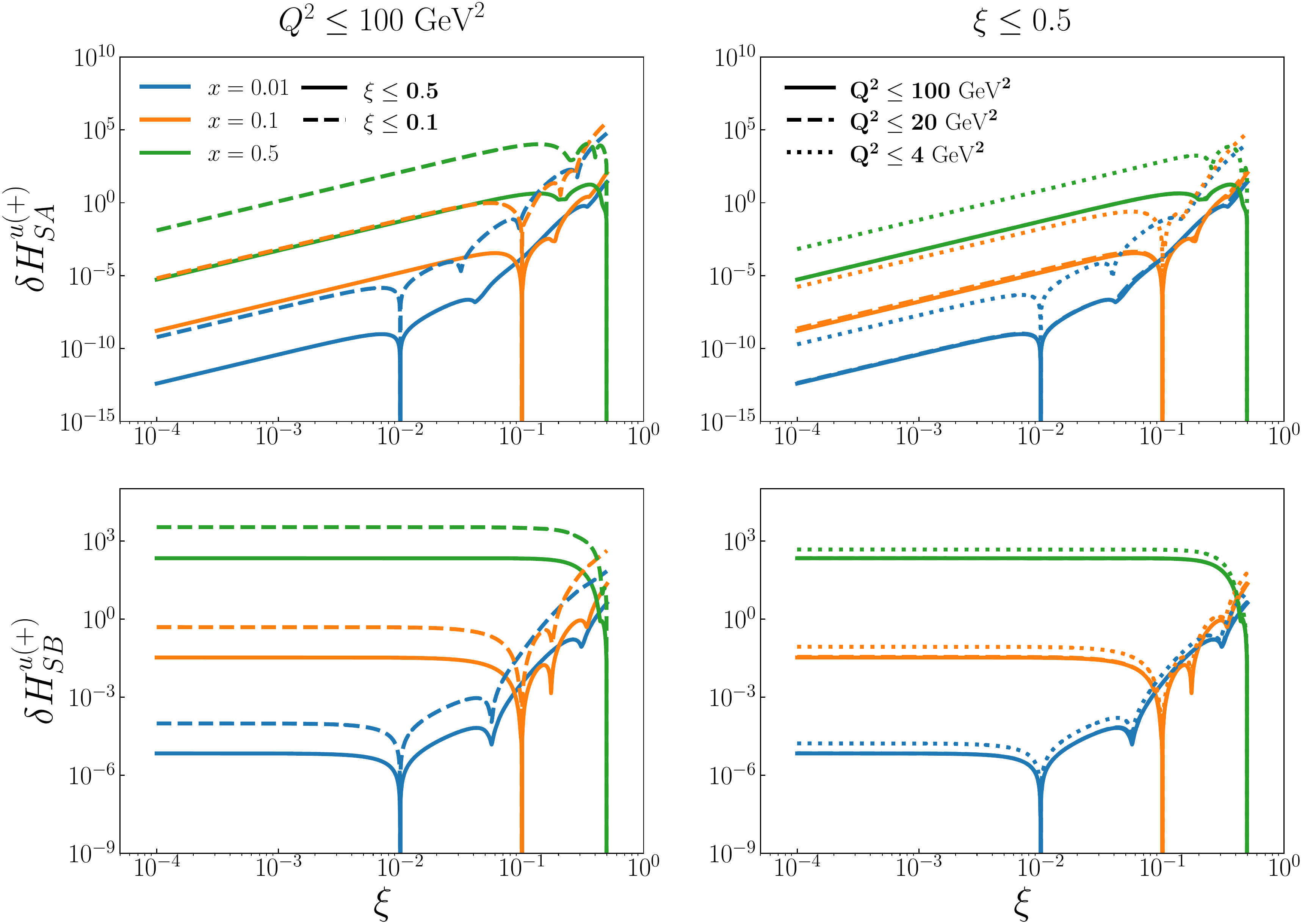}
\caption{
{\it Top row:} Results for $\delta{H}_S$ from Eq.~\eqref{eq:error} for type A SGPDs as a function of $\xi$, for three values of $x=0.01,\,0.1,\,0.5$. 
{\it Bottom row:} Results for $\delta{H}_S$ from Eq.~\eqref{eq:error} for type B SGPDs as a function of $\xi$, for three values of $x=0.01,\,0.1,\,0.5$.
In both cases, the left panels indicate results for cuts in the simulated CFF data in $\xi$, and the results in the right panels illustrate constraints on the SGPDs associated with cuts in $Q^2$ on the simulated CFF data. In each row the solid lines are the same and represent contraints from the full simulated data set.
}
\label{f.NLOHMC2}
\end{figure}

The results presented in Fig.~\ref{f.NLOHMC2} further illustrate these observations, where the top row corresponds to constraints on $\delta{H}_{S}$ for type A SGPDs and the bottom row corresponds to constraints on $\delta{H}_{S}$ for type B SGPDs. In each case, various cuts are made to the simulated data to illustrate the impact of changes in the $\xi$ and $Q^2$ range of the simulated data. In all four plots, results are shown as a function of $\xi$ for three values of $x=0.01,\,0.1,\,0.5$. In the left two panels, the entire $Q^2$ range of simulated data is included, the solid lines then show the constraint on $\delta{H}_{S}$ when the entire $\xi$ range of the simulated data is also included, and the dashed lines illustrate the case when a cut is made on the simulated data so that $\xi \leqslant 0.1$. We see that for type A SGPDs the large $\xi$ data constrains $\delta{H}_{S}$ by several orders of magnitude relative to only having data for $\xi \leqslant 0.1$ (difference between solid and dashed lines), and this remains true for all $x$ in the studied domain $x \in [0.01,\,0.5]$. Note that by construction $\delta{H}_S = 0$ when $x=\xi$, which explains the dips in $\delta{H}_S$ at these points. In the bottom left panel we see that large $\xi$ data also provides important constraints on type B SGPDs, however, not to the extent as for type A, and the dependence on $\xi$ is mild. These findings reflect the features of the SGPDs illustrated in the bottom row of Fig.~\ref{f.NLOHS}.

The right panels in Fig.~\ref{f.NLOHMC2} illustrate how $\delta{H}_S$ is impacted by varying the range of $Q^2$ spanned by the simulated CFF data. The solid lines in these plots are identical results to those in the left panels, however, in this case the dashed and dotted lines represent the results when the simulated CFF data is restricted to $Q^2 \leqslant 20\,$GeV$^2$ and $Q^2 \leqslant 4\,$GeV$^2$, respectively.  Notice that in the top right panel of \fref{NLOHMC2} the solid and dashed lines are fairly close together indicating the additional data between 20 GeV$^2$ and 100 GeV$^2$ does not provide substantially more constraint on the SGPDs than was already gained from the $Q^2\leqslant 20$ GeV$^2$ data.  In the bottom right panel of \fref{NLOHMC2}, the results do not seem to change significantly even from $Q^2 \leqslant 4$ GeV$^2$ to $Q^2 \leqslant 20$ GeV$^2$.  Therefore, while some lever arm in $Q^2$ is necessary for evolution to constrain the SGPDs, having a large lever arm in $Q^2$ does not seem as important as having data with a large lever arm in $\xi$. The results presented in Fig.~\ref{f.NLOHMC2} reflect the various attributes of the type A and type B SGPDs, and are independent of the choice of proxy model for the ``true'' GPDs. Our results for the SGPDs associated with the $E^{u(+)}$ GPDs which are of type B are similar to those in the bottom row of Fig.~\ref{f.NLOHMC2}.

Going back to Figs.~\ref{f.NLOsgpds} and \ref{f.NLOEsgpds}, we have illustrated as the hashed areas the regions for the SGPDs that would be excluded by the need to satisfy positivity constraints on the $H^{u(+)}$ and $E^{u(+)}$ GPDs, which read~\cite{Pobylitsa:2001nt,Pobylitsa:2002gw,Kirch:2005in}
\begin{align}
\label{eq:Hpos2}
\left|H^q(x,\xi,t;\mu^2)\right| &\leqslant 
\sqrt{\left(1 - \frac{t_{\rm min}\,\xi^2}{t_{\rm min} - t}\right)
\frac{q(x_{\rm in};\mu^2)\,q(x_{\rm out};\mu^2)}{1-\xi^2}}, &
\left|E^q(x,\xi,t;\mu^2)\right| &\leqslant \frac{2M}{\sqrt{t_{\rm min} - t}}
\sqrt{q(x_{\rm in};\mu^2)\,q(x_{\rm out};\mu^2)}.
\end{align}
It is clear that these positivity constraints can place significant constraints on the SGPDs, and could therefore play an important role in a global analysis of DVCS data to infer the GPDs. While this is an attractive method of constraining the SGPDs, it is important to note that these inequalities can be violated by regularization and renormalization effects in QCD~\cite{Collins:2021vke}.

\section{SHADOW GPDS AND PHENOMENOLOGY}
\label{s.sgpdimpact}
A driving motivation for DVCS experiments and the extraction of GPDs from this data is to determine key properties of the nucleon. An important example is the quark and gluon total angular momentum, which is related to the second moments of the helicity-independent GPDs by the Ji sum rule~\cite{Ji:1996ek}:
\begin{align}
J^a(\mu^2) &= \frac{1}{2}\int_{-1}^1 \dd x\ x\lf[H^a(x,0,0;\mu^2) + E^a(x,0,0;\mu^2)\rg]
= \frac{1}{2}\lf[A^a(0,\mu^2) + B^a(0,\mu^2)\rg],
\label{e.Jisum}
\end{align}
where $\frac{1}{2} = \sum_a\,J^a(\mu^2)$, and $A^a(t,\mu^2)$ and $B^a(t,\mu^2)$ are the gravitational form factors defined in Eqs.~\eqref{eq:emtffs} and \eqref{eq:emtcurrent}. In light of the discussion in the previous section, where simulated CFF data with 1\% errors were used to put constraints on the SGPDs, we can take these results and infer the impact of SGPDs on determining the $u$-quark contribution to the Ji sum rule. The SGPDs for $H^a(x,0,0;\mu^2)$ vanish in the forward limit and therefore do not contribute to the Ji sum rule, where $A^a(0,\mu^2)$ is completely determined by the parton momentum fractions which are obtained from the collinear PDFs. However, the SGPDs for $E^a(x,0,0;\mu^2)$ are of type B and do not necessarily vanish in the forward limit and will impact the Ji sum rule. Taking our results for the true and shadow GPDs we find that $A^u(0,\mu_0^2) = 0.389$ and $B^u(0,\mu_0^2) = 0.219 \pm 0.009$, where the momentum fraction has no error from the SGPDs because they vanish in the forward limit, while the error in $B^a(0,\mu_0^2)$ is associated with the contribution of $\delta E_S$ [see Eq.~\eqref{eq:error}] to the second moment of the $E^a$ GPDs from the type B SGPDs.
We therefore find that with accurate data over a wide range of $\xi$ and $Q^2$ it is possible to sufficiently constrain the SGPDs for an accurate determination of the Ji sum rule.

The pressure and shear force distributions within the nucleon are also of great interest. These mechanical properties of the nucleon can be obtained from the $C^a(t,\mu^2)$ gravitational form factors which are associated with the $\xi^2$ term from the second moment of both the $H^a$ and $E^a$ helicity-independent GPDs [see Eq.~\eqref{eq:emtffs}]. As discussed in Sec.~\ref{sec:proxy} and App.~\ref{ss.CFFs}, these form factors come entirely from the D-term of the GPDs. The D-term can be uniquely extracted from the subtraction constant in the dispersion relation [see Eq.~\eqref{e.disp}]~\cite{Dutrieux:2021nlz} which implies that a SGPD associated with the D-term can be completely constrained. Therefore SGPDs should not contribute to the pressure and shear stress distributions.

The GPDs provide a 3-dimensional spatial tomography of the nucleon, which is obtained by taking the 2-dimensional Fourier transform in the transverse plane of the GPDs at zero skewness $(\xi=0)$. This procedure gives the impact-parameter dependent PDFs~\cite{Burkardt:2000za}. To perform this Fourier transform accurate knowledge of the $t$ dependence of the GPDs is required at $\xi=0$. However, $t$-dependent data at $\xi=0$ is not experimentally accessible, so one has to extract the $t$ dependence from data at nonzero skewness and extrapolate to $\xi=0$. The type A SGPDs vanish when $\xi=0$, and our analysis in Sec.~\ref{s.sgpdev} also shows that these SGPDs can be very well constrained at small $\xi$. Therefore, the impact of type A SGPD on nucleon tomography would be minimal if CFF data over a large $\xi$ and $Q^2$ range can be obtained. The type B SGPDs can also be well constrained for small $\xi$ and small $x$ (see bottom panels of Fig.~\ref{f.NLOHMC2}). On the other hand, at higher $x$ these type B SGPDs are not well constrained, and their impact on nucleon tomography could be substantial. This scenario will need to be explored in a more complete global analysis of real and simulated CFFs. In addition, it is necessary to explore SGPDs whose $x$, $\xi$, and $t$ dependence is not separable. If the $t$ dependence of these SGPDs in the limit $\xi \to 0$ is substantially different from that of the true GPD, they would have a significant impact on nucleon tomography. A quantitative analysis of the potential impact of SGPDs on tomography requires a thorough exploration of the possible $t$-dependencies of SGPDs.

\section{CONCLUSION}
\label{s.conclusion}
We have explored the ability of QCD evolution to help limit the size of some example SGPDs when constrained by simulated CFF data over a large range in $\xi$ and $Q^2$. This was achieved by comparing the sample CFFs obtained from a Monte Carlo sampling of linear combinations of three example SGPDs, to the simulated CFFs obtained from VGG model GPDs that are used as a proxy for the ``true'' GPDs.  We separately conduct this analysis for both type A and type B SGPDs, where for type A SGPDs the forward limit constraint is imposed via the skewness dependence and the $t$ dependence of type B SGPDs ensures the correct forward limit. This study demonstrates that evolution can provide important constraints on at least the class of SGPDs represented by this sampling. Specifically, for type A SGPDs having data at large skewness leads to these SGPDs being very well constrained at smaller values of $\xi$ over the full range of $x$. We also find that such a level of constraint for type A SGPDs is not reliant on a very large lever arm in $Q^2$, with the data up to $Q^2 \simeq 20\,$GeV$^2$ proving to be sufficient. For type B SGPDs we find that having data at larger values of $\xi$ only leads to these SGPDs being well constrained in the region where both $\xi$ and $x$ are small.

These findings are independent of the proxy model used for the ``true'' GPDs because the results are driven by the skewness dependence of the SGPDs not the features of the proxy model. However, it is important to note that we have only conducted an analysis of a small representation of all possible SGPDs. While we have seen the potential of QCD evolution to constrain the class of SGPD under study, we currently cannot conclude that this is true for all SGPDs. We have therefore demonstrated that QCD evolution is a necessary but potentially not a sufficient means of constraining SGPD. A path forward would be to use more flexible parametrizations for the GPDs, coupled with simulated CFF data over a large range in $\xi$ and $Q^2$, and perform closure tests that provide robust uncertainty quantification.

We also explored the potential impact SGPDs could have on extracting information on key physical properties of the nucleon, such as the total quark and gluon angular momentum, and the pressure and shear force distributions. We find that shadow $H$ GPDs are not relevant to the total angular momentum because of the forward limit constraint. However, shadow $E$ GPDs could lead to significant contributions which depend on the uncertainty of the data and kinematics covered. SGPDs would not contribute to the determination of pressure and shear force distributions, because it is argued that the D-term can be uniquely determined from the subtraction constant in the CFF dispersion relation. We find the type A SGPDs would not have a significant impact on nucleon spatial tomography provided that data is available at large enough $\xi$ to impose adequate constraint in the small $\xi$ region. This remains true for the type B SGPDs in the region of small $x$. But the lack of constraint at small $\xi$ and large $x$ would lead to a significant impact on tomography in that region. A more quantitative analysis of the potential impact of SGPDs on tomography would require a detailed exploration of the possible $t$ dependence of SGPDs.

\begin{acknowledgments}
IC and EM were supported by the U.S.~Department of Energy, Office of Science, Office of Nuclear Physics, contract no.~DE-AC02-06CH11357.
LG was supported in part by the  Office of Nuclear Physics under contract No.~DE-FG02-07ER41460.	
AF was supported by the U.S. Department of Energy Office of Science, Office of Nuclear Physics under Award Number DE-FG02-97ER-41014.
AM was supported in part by the National Science Foundation under Grant  No.~PHY-2110472.
AP was supported in part by the National Science Foundation under Grant  No.~PHY-2012002.
The work of AP, WM, and NS was supported in part by the U.S. Department of Energy Contract No.~DE-AC05-06OR23177, under which Jefferson Science Associates, LLC operates Jefferson Lab. The work of NS was supported by the DOE, Office of Science, Office of Nuclear Physics in the Early Career Program. This research is also supported in part within the framework of the Quark-Gluon Tomography (QGT) Topical Collaboration, under contract no.~DE-SC0023646.
\end{acknowledgments}

\appendix
\section{COMPTON FORM FACTORS, HARD-SCATTERING COEFFICIENTS, AND THE D-TERM}
\label{ss.CFFs}
When QCD factorization theorems apply, CFFs can be expressed as convolutions of hard-scattering coefficient functions with the GPDs, as expressed in \eref{gpd2cff}. At leading order (LO) in $\alpha_s$, the coefficient functions associated with the helicity-independent ($H^a,\,E^a$) and helicity-dependent ($\tilde{H}^a,\,\tilde{E}^a$) GPDs are given by~\cite{Moutarde:2013qs}
\begin{align}
C^q_0(x,\xi,Q^2,\mu^2) &= -e_q^2
\left(\frac{1}{\xi-x+i\varepsilon} - \frac{1}{\xi+x-i\varepsilon}\right), &
C^g_0(x,\xi,Q^2,\mu^2) &= 0, \\
\tilde{C}^q_0(x,\xi,Q^2,\mu^2) &= -e_q^2
\left(\frac{1}{\xi-x+i\varepsilon} + \frac{1}{\xi+x-i\varepsilon}\right), &
\tilde{C}^g_0(x,\xi,Q^2,\mu^2) &= 0.
\label{e.LOcoeffs}
\end{align}
where $e_q$ is the quark electric charge and we see that the gluon GPD does not contribute to the CFFs at LO. The next-to-leading order (NLO) coefficient functions read~\cite{Ji:1997nk, Mankiewicz:1997bk, Belitsky:1999sg, Freund:2001rk, Freund:2001hd, Pire:2011st, Moutarde:2013qs} 
\begin{align} 
C^q_1(x,\xi,Q^2,\mu^2) &= \frac{\alpha_s\,e_q^2\,C_F}{4\pi}
\lf[c_1^q(x,\xi) - c_1^q(-x,\xi)\rg], &
C^g_1(x,\xi,Q^2,\mu^2) &= \frac{\alpha_s\,\sum e_q^2\,T_F}{4\pi}
\lf[c_1^g(x,\xi) - c_1^g(-x,\xi)\rg], \\
\tilde{C}^q_1(x,\xi,Q^2,\mu^2) &= \frac{\alpha_s\,e_q^2\,C_F}{4\pi}
\lf[\tilde{c}_1^q(x,\xi) + \tilde{c}_1^q(-x,\xi)\rg], &
\tilde{C}^g_1(x,\xi,Q^2,\mu^2) &= \frac{\alpha_s\,\sum e_q^2\,T_F}{4\pi}
\lf[\tilde{c}_1^g(x,\xi) + \tilde{c}_1^g(-x,\xi)\rg],
\label{e.NLOcoeffs}
\end{align}
where
\begin{align}
c_1^q &= \frac{1}{x+\xi-i\varepsilon}\left[9-3\frac{x+\xi}{x-\xi}\ln\left(\frac{x+\xi}{2\xi}-i\varepsilon\right)-\ln^2\left(\frac{x+\xi}{2\xi}-i\varepsilon\right) - \ln\left(\frac{Q^2}{\mu^2}\right)\left(3+2\ln\left(\frac{x+\xi}{2\xi}-i\varepsilon\right)\right)\right], \allowdisplaybreaks \\
\tilde{c}_1^q &= \frac{1}{x+\xi-i\varepsilon}\left[9-\ph{3}\frac{x+\xi}{x-\xi}\ln\left(\frac{x+\xi}{2\xi}-i\varepsilon\right)-\ln^2\left(\frac{x+\xi}{2\xi}-i\varepsilon\right) - \ln\left(\frac{Q^2}{\mu^2}\right)\left(3+2\ln\left(\frac{x+\xi}{2\xi}-i\varepsilon\right)\right)\right], \\
c_1^g &= \frac{x}{(x+\xi-i\varepsilon)(x-\xi+i\varepsilon)}
\left[\ph{-}2\frac{x+3\xi}{x-\xi}\ln\left(\frac{x+\xi}{2\xi}-i\varepsilon\right)
-\frac{x+\xi}{x-\xi}\ln^2\left(\frac{x+\xi}{2\xi}-i\varepsilon\right)
- \ln\left(\frac{Q^2}{\mu^2}\right)2\frac{x+\xi}{x-\xi}\ln\left(\frac{x+\xi}{2\xi}-i\varepsilon\right)\right], \\
\tilde{c}_1^g &= \frac{x}{(x+\xi-i\varepsilon)(x-\xi+i\varepsilon)}
\left[-2\frac{3x+\xi}{x-\xi}\ln\left(\frac{x+\xi}{2\xi}-i\varepsilon\right)
+ \frac{x+\xi}{x-\xi}\ln^2\left(\frac{x+\xi}{2\xi}-i\varepsilon\right)
+ \ln\left(\frac{Q^2}{\mu^2}\right)2\frac{x+\xi}{x-\xi}\ln\left(\frac{x+\xi}{2\xi}-i\varepsilon\right)\right].
\end{align}
We see that the $C^a$ are anti-symmetric in $x$, and therefore only the anti-symmetric quark singlet and gluon combinations of $H^a$ and $E^a$ give nonzero contributions to the CFFs. Similarly, the $\tilde{C}^a$ are symmetric in $x$ so that only the symmetric quark non-singlet and gluon combinations of $\tilde{H}^a$ and $\tilde{E}^a$ give nonzero contributions to the associated CFFs.

CFFs are complex-valued functions where the real and imaginary parts obey dispersion relations \cite{Anikin:2007yh,Diehl:2007jb}. A LO example reads
\begin{equation}
\label{e.disp}
{\rm Re}\mathcal{F}(\xi,t,Q^2) 
= \int_{-1}^1\frac{\dd{\xi'}}{\pi}\ {\rm Im}\mathcal{F}(\xi',t,Q^2)
\left(\frac{1}{\xi - \xi'}-\frac{1}{\xi + \xi'}\right) \pm S(t,Q^2),
\end{equation}
where $\mathcal{F} = \mathcal{H},\,\mathcal{E}$, and the plus sign is for the CFF $\mathcal{H}$. Therefore, the subtraction constant $S(t,Q^2)$ is the same for $\mathcal{H}$ and $\mathcal{E}$ up to an overall sign. This real-valued subtraction constant is related to the D-term contribution to the GPDs:
\begin{equation}
\label{e.subconst}
S(t,Q^2) = 2\int_{-1}^1\dd\alpha\frac{D(\alpha,t,Q^2)}{1-\alpha}.
\end{equation}
This dispersion relation implies that the D-term can be directly determined from CFF data, which justifies excluding the D-term in our SGPDs.

\section{GPD POLYNOMIALITY RELATIONS\label{app:gpds}}
A remarkable property of GPDs is known as polynomiality, which means that the Mellin moments of the GPDs are even polynomials in $\xi$. For the nucleon GPDs, polynomiality states that moments of the quark and gluon GPDs satisfy~\cite{Diehl:2003ny}
\begin{subequations}
\label{e.poly}
\begin{align}
\label{eq:Hpoly}
\int_{-1}^1\dd{x}\,x^sH^a(x,\xi,t;\mu^2) &= 
\sum_{i=0\,(\text{even})}^{s}\ (2\xi)^{i}\,A^a_{s+1,i}(t,\mu^2) 
+ {\rm mod}(s,2)\,(2\xi)^{s+1}\,C^a_{s+1}(t,\mu^2),\\
\label{eq:Epoly}
\int_{-1}^1\dd{x}\,x^sE^a(x,\xi,t;\mu^2) &= 
\sum_{i=0\,(\text{even})}^{s}\ (2\xi)^{i}\,B^a_{s+1,i}(t,\mu^2) 
- {\rm mod}(s,2)\,(2\xi)^{s+1}\,C^a_{s+1}(t,\mu^2),\\
\int_{-1}^1\dd{x}\,x^s\tilde{H}^a(x,\xi,t;\mu^2) &= 
\sum_{i=0\,(\text{even})}^{s}\ (2\xi)^{i}\,\tilde{A}^a_{s+1,i}(t,\mu^2),\\
\int_{-1}^1\dd{x}\,x^s\tilde{E}^a(x,\xi,t;\mu^2) &= 
\sum_{i=0\,(\text{even})}^{s}\ (2\xi)^{i}\,\tilde{B}^a_{s+1,i}(t,\mu^2),
\end{align}
\end{subequations}
where mod($s,2$) gives $1$ if $s$ is odd and zero otherwise. Note, because we use the Ji definition of the gluon GPDs, the polynomiality relations for the quarks and gluons take the same form. The quark and gluon total angular momentum is related to the second moments of the GPDs by the Ji sum rule~\cite{Ji:1996ek}:
\begin{align}
J^a(\mu^2) &= \frac{1}{2}\lf[A^a_{20}(0,\mu^2)+B^a_{20}(0,\mu^2)\rg] 
= \frac{1}{2}\int_{-1}^1 \dd x\ x\lf[H^a(x,0,0;\mu^2)+E^a(x,0,0;\mu^2)\rg],
\label{e.Jisum}
\end{align}
where $A^a_{20}$ and $B^a_{20}$ are the $s=1,i=0$ terms from Eqs.~\eqref{eq:Hpoly} and \eqref{eq:Epoly}. For completeness, we include the definition for the leading-twist quark and gluon helicity-dependent GPDs for a spin-half target:
\begin{align}
\label{e.DqGPD}
P\cdot n\int\frac{\dd \lambda}{2\pi}\ e^{ix P\cdot n\lambda}
\left< p'\lf|\bar{\psi}^q(-\tfrac12 \lambda n)\,\sh{n}\gamma_5\, \psi^q (\tfrac12 \lambda n)\rg|p\right>
&= 
\bar{u}(p')
\left[\ph{x\,}\tilde{H}^q(x,\xi,t;\mu^2)\,\sh{n}\gamma_5 + 
\ph{x}\tilde{E}^q(x,\xi,t;\mu^2)\,\frac{\Delta\cdot n\, \gamma_5}{2M}
\right]u(p), \\
\label{e.DgGPD}
-i\,n_\mu n_\nu \int\frac{\dd \lambda}{2\pi}\ e^{ix P\cdot n\lambda}
\big< p'\lf|G^{\mu\alpha}(-\tfrac12 \lambda n)\,\tilde{G}_\alpha{}^\nu(\tfrac12 \lambda n)\rg|p \big>
&= \bar{u}(p') 
\left[x\,\tilde{H}^g(x,\xi,t;\mu^2)\, \sh{n}\gamma_5
+ x\,\tilde{E}^g(x,\xi,t;\mu^2)\, \frac{\Delta\cdot n\, \gamma_5}{2M} 
\right]u(p),
\end{align}
where the gluon dual field strength tensor is given by $\tilde{G}^{\mu\nu} = \frac{1}{2}\varepsilon^{\mu\nu\alpha\beta}G_{\alpha\beta}$. We again use the Ji convention~\cite{Ji:1998pc} for the gluon GPDs, which differs from the Diehl convention~\cite{Diehl:2003ny} by $\tilde{F}^g_{\rm Diehl} = 2\,x\,\tilde{F}^g_{\rm Ji}$.

\section{QCD EVOLUTION OF GPDS}
\label{s.evolution}
We compute the LO evolution of the helicity-independent GPDs following Ref.~\cite{Ji:1996nm}. For the quark non-singlet GPDs $F^q_{\rm NS} = F^q(x,\xi,t) + F^q(-x,\xi,t)$, when $x\geqslant\xi$ the evolution is given by
\begin{align}
  \frac{\dd{F^q_{\rm NS}}(x,\xi,t;\mu^2)}{\dd\ln{\mu^2}}
  =&\ \frac{\alpha_s(\mu^2)}{2\pi} \left[ \left(\frac{3}{2}+\int_\xi^x\frac{\dd{y}}{y-x}+\int_{-\xi}^x\frac{\dd{y}}{y-x}\right)C_F\,F^q_{\rm NS}(x,\xi,t;\mu^2)
+ \int_x^1\frac{\dd{y}}{y}\ 
P_{qq}\lf(\frac{x}{y},\frac{\xi}{y}\rg)
F^q_{\rm NS}(y,\xi,t;\mu^2)\right],
\label{e.dglapns}
\end{align}
where $F=H,\,E$, $C_F=4/3$, and the splitting function $P_{qq}$  at LO is given by
\begin{equation}
    P_{qq}(x,\xi)=C_F\frac{x^2+1-2\xi^2}{(1-x)(1-\xi^2)}.
    \label{e.Pqqdglap}
\end{equation}
In the region $x \leqslant -\xi$, the evolution is given by \eref{dglapns} with the replacement of $\int_x^1 \to -\int_{-1}^x$. Finally, in the region $-\xi \leqslant x \leqslant \xi$ the non-singlet evolution equation is
\begin{align}
  \frac{\dd{F^q_{\rm NS}}(x,\xi,t;\mu^2)}{\dd\ln{\mu^2}}=&\frac{\alpha_s(\mu^2)}{2\pi} \left[ \left(\frac32 + \int_\xi^x\frac{\dd{y}}{y-x}+\int_{-\xi}^x\frac{\dd{y}}{y-x}\right)C_F\,F^q_{\rm NS}(x,\xi,t;\mu^2)\right. \notag \\
&\hspace{50mm}
\left. + \lf[\int_x^1\frac{\dd{y}}{y}\ P'_{qq}\lf(\frac{x}{y},\frac{\xi}{y}\rg) - \int_{-1}^x\frac{\dd{y}}{y}\ P'_{qq}\lf(\frac{x}{y},-\frac{\xi}{y}\rg)\rg] 
F^q_{\rm NS}(y,\xi,t;\mu^2)\right],
  \label{e.erblns}
\end{align}
where
\begin{equation}
    P'_{qq}(x,\xi)=C_F\frac{x+\xi}{\xi(1+\xi)}\left(1+\frac{\xi}{1-x}\right).
    \label{e.Pqqerbl}
\end{equation}
The quark singlet $[F^{\rm S} = \sum_q F^{q(+)}$, where  $F^{q(+)} = F^q(x,\xi,t) - F^q(-x,\xi,t)]$ and gluon GPDs mix under evolution, and when $x \geqslant \xi$ their evolution is given by
\begin{align}
  \frac{\dd{F^{\rm S}}(x,\xi,t;\mu^2)}{\dd\ln{\mu^2}}=&\frac{\alpha_s(\mu^2)}{2\pi} \left[\left(\frac{3}{2}+\int_\xi^x\frac{\dd{y}}{y-x}+\int_{-\xi}^x\frac{\dd{y}}{y-x}\right)
C_F\,F^{\rm S}(x,\xi,t;\mu^2)\right. \notag \\
&\hspace{35mm}
\left.+\int_x^1\frac{\dd{y}}{y}\ P_{qq}\lf(\frac{x}{y},\frac{\xi}{y}\rg) F^{\rm S}(y,\xi,t;\mu^2) + 2\,n_f\int_x^1\frac{\dd{y}}{y}\ P_{qg}\lf(\frac{x}{y},\frac{\xi}{y}\rg) F^{\rm g}(y,\xi,t;\mu^2)\right], \label{e.dglaps} \\
\frac{\dd{F^{\rm g}}(x,\xi,t;\mu^2)}{\dd\ln{\mu^2}}=&\frac{\alpha_s(\mu^2)}{2\pi} \left[\left(\frac{11}{6}-\frac{n_f}{3C_A}+\int_\xi^x\frac{\dd{y}}{y-x}+\int_{-\xi}^x\frac{\dd{y}}{y-x}\right) C_A\,F^{\rm g}(x,\xi,t;\mu^2)\right. \notag \\
&\hspace{11mm}
+ \frac{1}{2} \int_x^1\frac{\dd{y}}{y}\ P_{gq}\lf(\frac{x}{y},\frac{\xi}{y}\rg)
  \lf[F^{\rm S}(y,\xi,t;\mu^2)-F^{\rm S}(-y,\xi,t;\mu^2)\rg]
\left.+\int_x^1\frac{\dd{y}}{y}\ P_{gg}\lf(\frac{x}{y},\frac{\xi}{y}\rg) 
F^{\rm g}(y,\xi,t;\mu^2)\right], \label{e.dglapg} 
\end{align}
with $n_f$ the number of active quark flavors and $C_A=3$. The additional splitting functions are
\begin{subequations}
\begin{align}
    &P_{qg}(x,\xi)=T_F\frac{x^2+(1-x)^2-\xi^2}{(1-\xi^2)^2}, \label{e.Pqgdglap} \\
    &P_{gq}(x,\xi)=C_F\frac{1+(1-x)^2-\xi^2}{x(1-\xi^2)}, \label{e.Pgqdglap} \\
    &P_{gg}(x,\xi)=C_A\frac{x^2-\xi^2}{x(1-\xi^2)^2}\left[1+\frac{2(1-x)(1+x^2)}{x^2-\xi^2}+\frac{1+x-2\xi^2}{1-x}\right], \label{e.Pggdglap}
\end{align}
\end{subequations}
where $T_F=1/2$. When $x \leqslant -\xi$ the quark singlet and gluon GPD evolution is given by the replacement of $\int_x^1 \to -\int_{-1}^x$ in \erefstwo{dglaps}{dglapg}. In the region $|x| \leqslant \xi$, the quark singlet and gluon GPD evolution equations are
\begin{align}
  \frac{\dd{F^{\rm S}}(x,\xi,t;\mu^2)}{\dd\ln{\mu^2}} &= \frac{\alpha_s(\mu^2)}{2\pi} \left[\left(\frac{3}{2}+\int_\xi^x\frac{\dd{y}}{y-x}+\int_{-\xi}^x\frac{\dd{y}}{y-x}\right)
C_F\,F^{\rm S}(x,\xi,t;\mu^2)\right. \notag \\
&
\hspace{30mm}
+ \lf[\int_x^1\frac{\dd{y}}{y}\ P'_{qq}\lf(\frac{x}{y},\frac{\xi}{y}\rg)
- \int_{-1}^x\frac{\dd{y}}{y}\ P'_{qq}\lf(\frac{x}{y},-\frac{\xi}{y}\rg)\rg] 
F^{\rm S}(y,\xi,t;\mu^2) \notag \\
&
\hspace{52mm}
\left. +\, 2\,n_f\left[\int_x^1\frac{\dd{y}}{y}\ P'_{qg}\lf(\frac{x}{y},\frac{\xi}{y}\rg)
-\int_{-1}^x\frac{\dd{y}}{y}\ P'_{qg}\lf(\frac{x}{y},-\frac{\xi}{y}\rg)\right] 
F^{\rm g}(y,\xi,t;\mu^2)\right], 
\label{e.erbls} \\
\frac{\dd{F^{\rm g}}(x,\xi,t;\mu^2)}{\dd\ln{\mu^2}} &= \frac{\alpha_s(\mu^2)}{2\pi} \left[\left(\frac{11}{6}-\frac{n_f}{3C_A}+\int_\xi^x\frac{\dd{y}}{y-x}+\int_{-\xi}^x\frac{\dd{y}}{y-x}\right) C_A\,F^{\rm g}(x,\xi,t;\mu^2)\right. \notag \\
&
\hspace{20mm}
+ \frac{1}{2} \lf[\int_x^1\frac{\dd{y}}{y}\ P'_{gq}\lf(\frac{x}{y},\frac{\xi}{y}\rg) 
- \int_{-1}^x\frac{\dd{y}}{y}\ P'_{gq}\lf(\frac{x}{y},-\frac{\xi}{y}\rg)\rg]
  \lf[F^{\rm S}(y,\xi,t;\mu^2)-F^{\rm S}(-y,\xi,t;\mu^2)\rg] \notag \\
&
\hspace{60mm}
\left. + \lf[\int_x^1\frac{\dd{y}}{y}\ P'_{gg}\lf(\frac{x}{y},\frac{\xi}{y}\rg) 
- \int_{-1}^x\frac{\dd{y}}{y}\ P'_{gg}\lf(\frac{x}{y},-\frac{\xi}{y}\rg)\rg] 
F^{\rm g}(y,\xi,t;\mu^2)\right], 
\label{e.erblg} 
\end{align}
where
\begin{subequations}
\begin{align}
    &P'_{qg}(x,\xi)=T_F\frac{(x+\xi)(1-2x+\xi)}{\xi(1+\xi)(1-\xi^2)}, \label{e.Pqgerbl} \\
    &P'_{gq}(x,\xi)=C_F\frac{(x+\xi)(2-x+\xi)}{x\xi(1+\xi)}, \label{e.Pgqerbl} \\
    &P'_{gg}(x,\xi)=-C_A\frac{x^2-\xi^2}{x\xi(1-\xi^2)}\left[1-\frac{\xi}{1-x}-\frac{2(1+x^2)}{(1+\xi)(x-\xi)}\right]. \label{e.Pggerbl}
\end{align}
\end{subequations}
To calculate the evolution in our analysis, we have implemented these expressions and utilized the technique developed by Vinnikov \cite{Vinnikov:2006xw} to make these calculations numerically more efficient.

\bibliographystyle{apsrev4-1}
\bibliography{bibliography}

\end{document}